\documentclass[articel,twocolumn,showpacs,amsmath,amssymb,floatfix]{revtex4-1}
\usepackage{graphicx}%Include figure files
\usepackage{dcolumn}%Align table columns on decimal point
\usepackage{bm}% bold math
\usepackage{amsmath}
\usepackage{float}
\begin{document}

\title{The Spiral Structure of the Milky Way Galaxy from Radio Observations
}

\author{Jawaher Alsobaie}
\affiliation{Department of Physics and Astronomy, University
of Delaware, Newark, DE 19716-2570, USA}

~\newline
\\
\\
\\

\begin{abstract}
The study of the Milky Way structure is vital importance for a better understanding of the universe. This study digs deeper into the structure of Milky Way Galaxy’s with emphasis to its spiral arms. Multiple data sources were sought including the NASA websites. As found out, the Milky Way is a disk shaped galaxy composed of four spiral arms, two large and the rest are smaller. The spiral arms are regions of actively forming new stars dominated by young stars, dust and gas. Dust and gas are primary materials for the formation of new stars.  This dust is also responsible for the reddish appearance of the stars since it absorbs more blue light than red. Besides the disk, there are numerous older stars, which appear white in color according to the picture by COBE. Additionally, the Milky Way contains a dark halo, extensive but void of luminous stars. The extensive halo was discovered due to the gravitational pull it exerts on existing visible matter.

\end{abstract}

\maketitle

\section{Introduction} \label{sec:intro}

The Milky Way is a collection of close to a billion stars bound by gravitational forces. The solar system is part of the Milky Way (Clements, 2015). As such sun is among the hundred billion stars that make the Milky Way. The desire to know more about the universe has always driven scientists into deeper research. A greater understanding of Milky Way, a galaxy in which the earth is positioned is not only satisfies the curious minds of millions of people around the world but it also amazing considering its interesting features. More importantly, events that occur around the galaxy may eventually determine the fate of the earth. A better understanding of major events and process may position scientist at a vantage position to predict and impeding danger such as collision with terrestrial bodies and save the situation if possible (Oort, Westerhout, 1958).  NASA has always searched the sky for potentially hazardous phenomena or object. In a recent undertaking NASA reactivated WISE, an Infrared survey, to scan the sky in 2013 after two years of hibernation in a mission to enable it to identify objects that may harm to the planet earth (NASA, 2015). The structure of the Milky Way, the galaxy in which the earth and the entire solar system is located, takes a central position in this study.
\begin{figure}
 \includegraphics[scale=0.38]{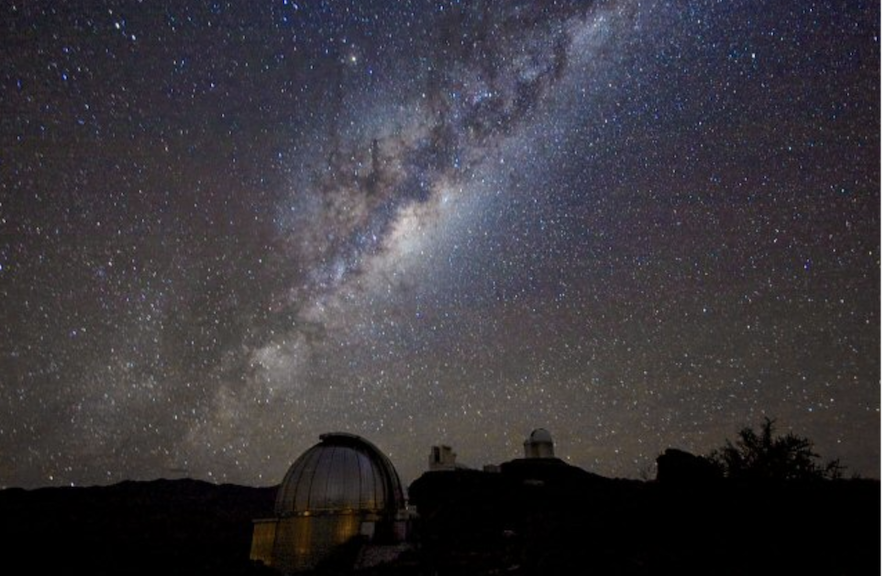}
\caption{The Milky Way appear as a band in the night sky. This picture was taken in the southern hemisphere, A. Fitzsimmons(ESO).}
\label{Figure 1}
\end{figure}

\section{Background Information }
The study of the galaxy dates back to more than 322 BC where most of the available information was based on speculation.  Aristotle noted propositions by Democritus and Anaxagoras who thought the galaxy was composed of distant stars. In contrast Aristotle thought that the Milky Way originated from exhalation of large stars being ignited in upper atmosphere (Kormendy, Bender, 2018). Although their views were largely criticized by other scholars, their views marked the start of a long journey of study that is yet to complete. Many more scholars after Aristotle publicized their findings about the galaxy based on observations (Oort, Westerhout, 1958). In 1048 Abū Rayhān, a Persian astronomer asserted that the Milky Way comprised countless nebula star-like fragments. Later, Avempace, an Andalusion astronomer came up with a new idea claiming the galaxy was made up of stars that were reflected in the sky to make them appear as a continuous mass of luminous object.
A more scientific approach to explore the universe took place in 1610 when Galileo Galilei observed the sky using a telescope. His approach proved that galaxy was in deed made of stars. Other ideas such as the galaxy rotate along with all the stars and other bodies and existence of multiple galaxies would later follow (Kormendy, Bender, 2018). 
Efforts to describe the shape of the galaxy took shape in 1785 when William Herschel attempted counting stars in various regions of the sky and came up with a diagram to describe the Milky Way’s shape. A more advanced telescope was developed in 1845 enabling Lord Rose to prove the existence of a spiral shaped nebulae and an elliptical one (Oort, Westerhout, 1958).
In 1920, an astronomical debate about the shape and size of the universe between Harlow Shapley and Heber D. Cutis was held in the National History Museum located in Washington DC. Curtis believed that the universe was composed of several galaxies citing that there were spiral nebulae outside the Milky Way while Shapley held that the universe had only one galaxy and the other phenomena were just gas clusters (Kormendy, Bender, 2018).
Although Curtis and Shapley’s debate ended in controversy, it sparked new efforts to pin down the nature of the universe. A few years later, Edwin Hubble brought the controversy to an end by proving that there were spiral nebulae external to the Milky Way and composed of stars rather than gas clusters (Kormendy, Bender, 2018). He used an innovated Hooker’s telescope to produce astronomical photographs with enough resolution to distinguish individual stars in the nebulae.
These approaches by early scientist to the study of the Milky Way were largely unsuccessful due to the limitations of the visible light that barely penetrates opaque matter.  In search for better methods to explore the galaxy, use of radio waves was adapted in the early 20th century starting with the use of the radio waves to communicate over long distance (Langford, 2013).
Later, an interference of radio communications in the Bell Telephone Company initiated an investigation by Karl Jansky to establish the source of extra wave signal. After investigation, Jansky concluded the source of extra radio waves was the Sagittarius in the Milky Way galaxy. Since then more study focused on use of radio waves to study the Milky Way (Lankford, 2013).

\section{Radio Observations}
\subsection{  21-cm Radio Line}
The prime drawback to optical mapping of the Galaxy is obscuration from interstellar dust. Radio observations do not have this handicap. It has solved this obstacle because radio waves can travel through the dust and gas and provide us with images. 

 Another major breakthrough in astronomy took place in the 1950s after the discovery of 21cm (Clements, 2015). The discovery of the 21 cm hydrogen radio in 1944 by H.C Halt was a big breakthrough in the study of galaxies.  Due to spin-spin interactions, the ground state of hydrogen is divided into two states. In the system formed, the energy of the system depends on whether the two spins are parallel or antiparallel. An antiparallel spin results into a lower energy in the system. The resulting difference in energy is equivalent to radiation of $21cm$ of $1420.405751786$ $MHz$ which is useful in radio astronomy (Kormendy, Bender, 2018).  With each transition of a hydrogen atom from a higher energy state to the lower one, a radio wave at 1420.405751786 is emitted.  Hydrogen is a major component of terrestrial bodies. The interstellar gas that gives rise to stars is mainly made up hydrogen element. In the whole milky way, close to $10\%$ by mass is comprises the interstellar gas, positioned in a thin disk that rotates around one position known as the Galactic Centre.  Since hydrogen is one of the most abundant elements in the universe and in the galaxy, there are a lot of line emissions from the galaxy. The line emissions were first observed by Purcell and Ewen in 1951.
The discovery by the two scientists has played a crucial role in the study of the universe. This is because astronomers have been able to measure the gas cloud movements using the spectral line in combination with Doppler Effect (Clements, 2015). Doppler Effect results from a frequency change of an emission when the emitting body or the observer moves. In particular, the study has facilitated the pinning down of the fact that the Milky Way is a spiral galaxy with two major arms as can be derived from images and spectrums captured by radio telescopes.
Taking images of the Milky Way remained a challenge for centuries since the earth is positioned in plane that is located in the Milky Way’s disk such that most parts of the galaxy are obstructed from earth’s sight.  According to NASA’s estimation, the earth is situated at a point two-thirds of the radius away from the galaxy’s center (NASA, 2015). A strategic position from which one can have a detailed view of the galaxy is a distant vertical location probably in another galaxy. The view would just be similar to that a person high above views the exact positions of objects in the ground. However, such a vantage position could be about 75000 light years away, an impossible distances to cover with the current spaceship (Oort, Westerhout, 1958).  

Today, scientists have been able to capture images of the galaxy using different radio waves. Powerful radio telescopes set at different wavelengths are use in this mapping (Kormendy, Bender, 2018).  Each wavelength brings out a perspective of the galaxy as shown below. In the image the galaxy’s disc is obscured in X-ray and optical images but visible in molecular and atomic hydrogen waves, as well as infra-red and gamma rays. 
\begin{figure}[h!]
 \includegraphics[scale=0.38]{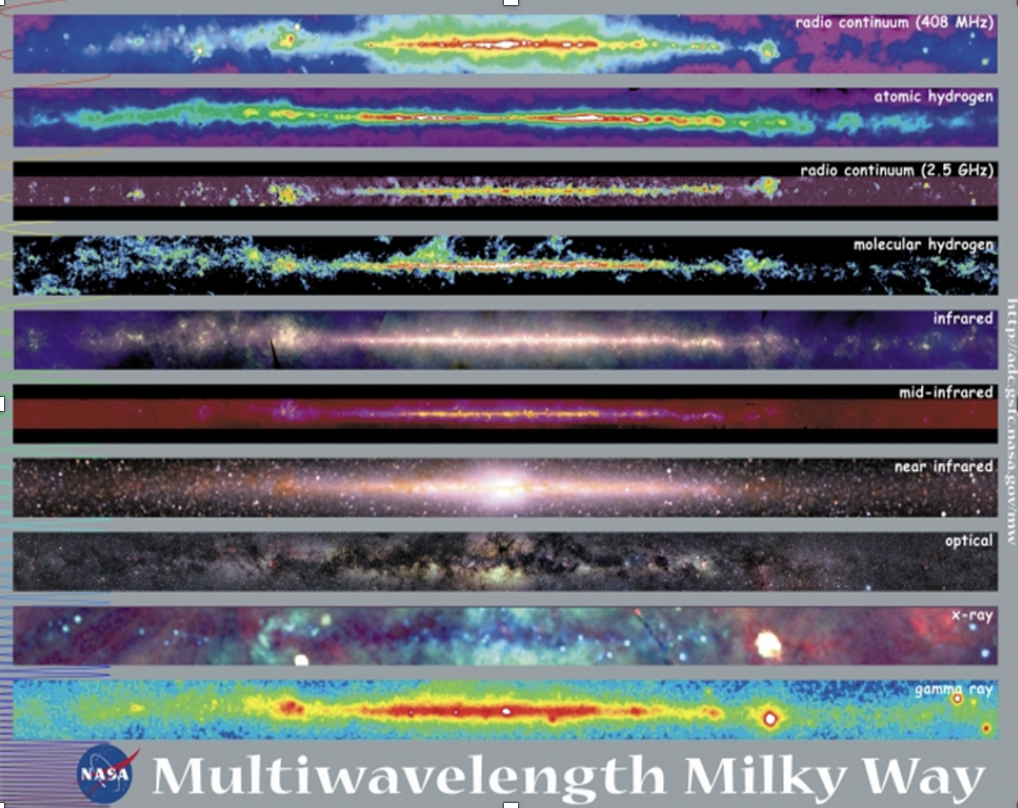}
\caption{Milk Way With Different Wavelengths(Clements, 2015).}
\label{Figure 2}
\end{figure}
\begin{figure}
 \includegraphics[scale=0.38]{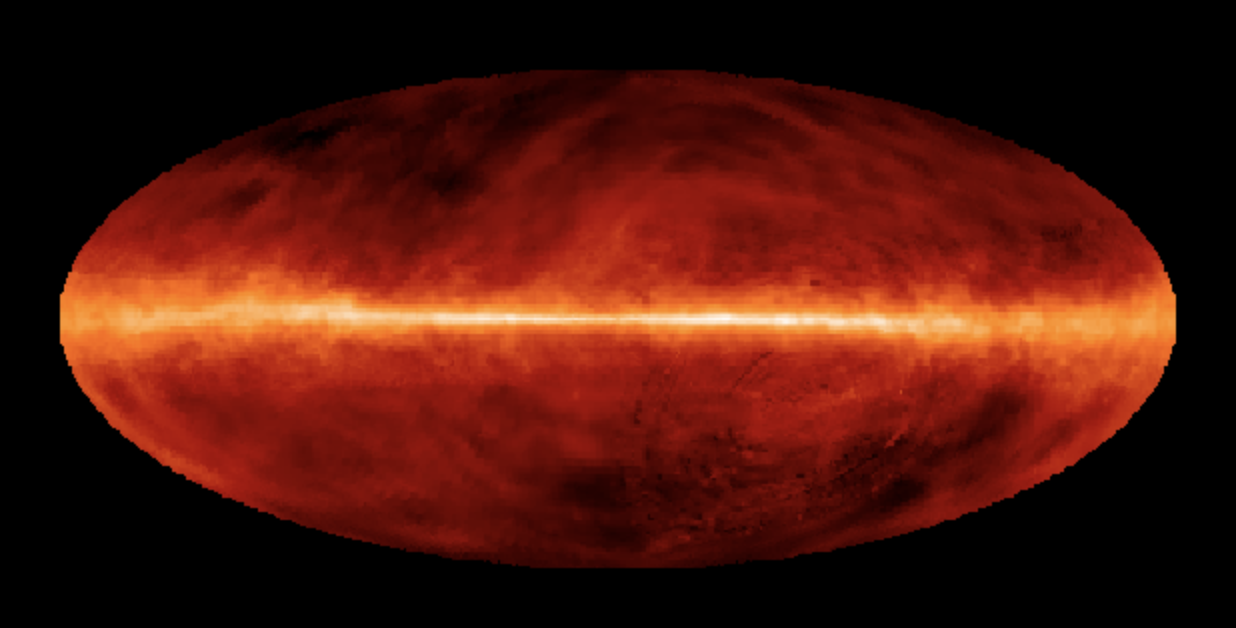}
\caption{A Sky Full Of Hydrogen,(J. Dickey (UMn), F. Lockman (NRAO), SkyView).}
\label{Figure 3}
\end{figure}

\subsection{  Results and the Experiments		
}
Experiments to reveal the map of the Milky Way take advantage of the hydrogen gas clouds. As discussed above, hydrogen gas has a natural emission of radio waves that make the 21 cm spectral line which is detectable from the earth’s surface. A radio telescope directed towards the Milky Way’s plane, radio waves emitted by the gas clouds can be obtained (Strasbourg, 2010).  
A telescope pointed in the direction of Galactic plane produce signals with which computer interface produce a spectrum like the one below. At around $1200$, power counts, there is receiver’s noise emission. Galactic emissions are observed at around $1420.4$ $MHz$ spread across $500$ $kHz$ with bumps which can be inferred to hydrogen emissions since $1420.406$ is its true frequency (Strasbourg, 2010). Further, the radial velocity of the group of hydrogen atoms responsible for this emission can be determined. As a result,  the data we receive from the telescope tells us how fast the gas is moving in the direction we are looking. 
 \begin{figure}[h!]
 \includegraphics[scale=0.38]{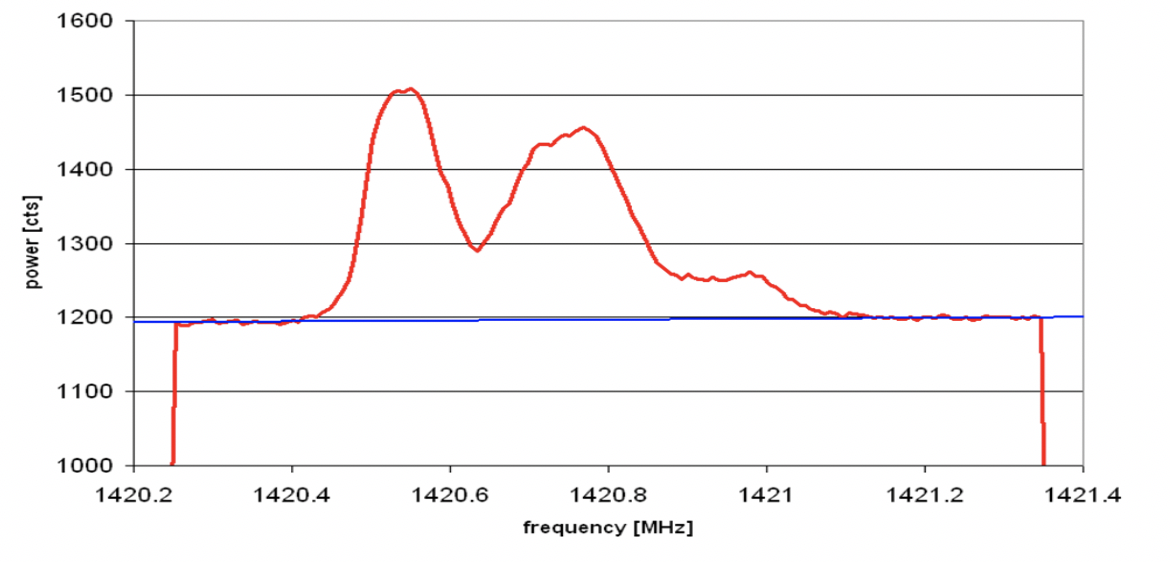}
\caption{ Milky Way Frequency, (Strasbourg, 2010).}
\label{Figure 4}
\end{figure}
 Ideally, the waves are supposed to be located at the true wavelength of the hydrogen gas that is $1420.4$ $MHz$ but the gas clouds from which the waves are emitted are in constant motion with some moving away and others towards the earth. Consequently, there are an emissions at a lower frequency from clouds moving away from the earth and emissions at higher frequency from gas clouds moving towards the earth. These differences allow estimation of the velocity the gases in the two directions (Kormendy, Bender, 2018). These measurements can in turn allow the determination of concentration of the gas in the Milky Way. If we can predict how fast different parts of the galaxy are moving as shown in figure 5, then we can work out where the gas must be concentrated along our line of sight to make the particular blend of radio waves which we observe. A lot of gas all moving at one speed for instance indicates a denser region of orbiting gas, possibly as part of a spiral arm(see figure 6)!
\begin{figure}
 \includegraphics[scale=0.38]{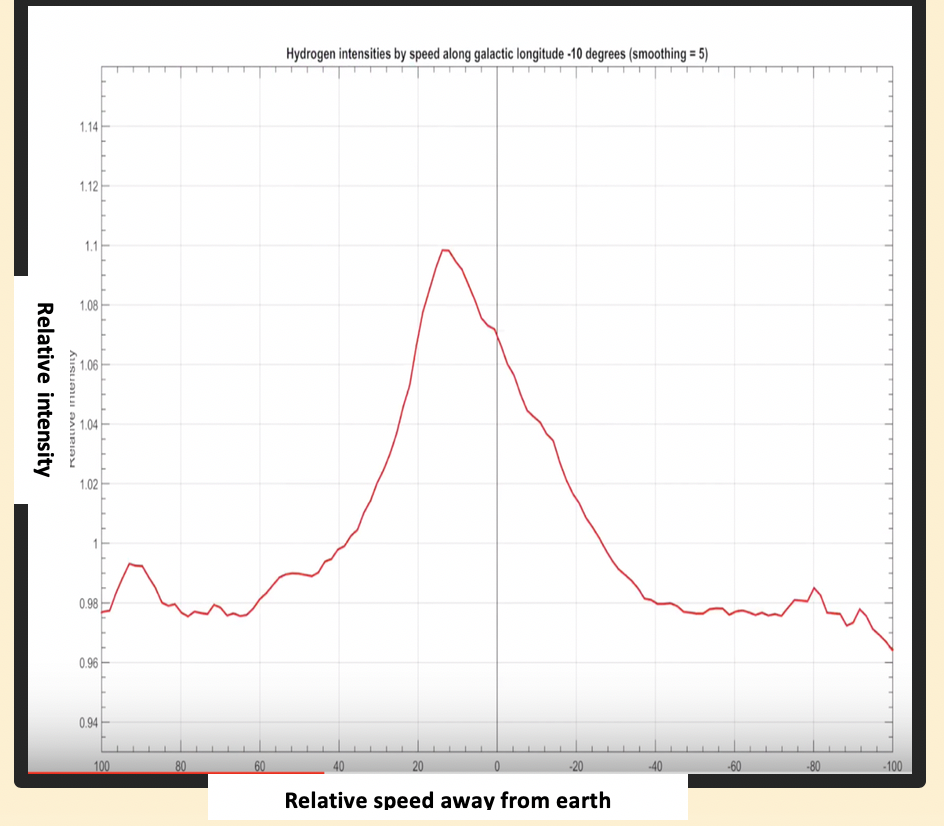}
\caption{ Relative intensity vs Relative speed,(Ben-David, 2015).}
\label{Figure 5}
\end{figure}
 \begin{figure}
 \includegraphics[scale=0.38]{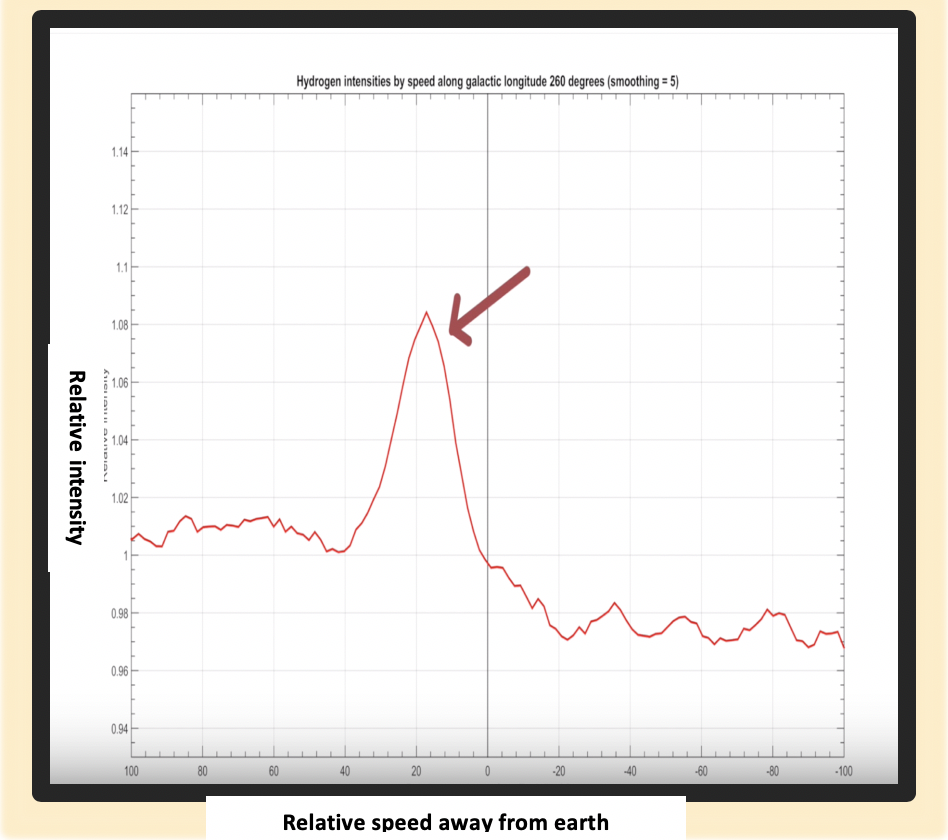}
\caption{ Part of a Spiral Arm,(Ben-David, 2015).}
\label{Figure 6}
\end{figure}
 
Using trigonometric laws, it is possible to plot graphs of hydrogen gases in specific lines of the disk (Ben-David, 2015). Repeating this process with multiple areas of the disk reveals the structure of the Milky Way. From such an experiment, have a disk with spiral region of varying densities as shown below. These regions represent the spiral arms of the Milky Way galaxy. There are three main visible sections of the spiral arm, a shorter arm referred to as Sagittarius and a longer one known as Perseus arm also the more distant piece is known as the Outer arm of the Milky Way as can we see in figure 7. What is even better it is not the first to find a map of the galaxy this way.This approach was first used by Kerr and Oort and Westerhout in 1950s. If we compare the resent result map to the old one, we find that they agree on the locations of several important features.(see figure 8)
\begin{figure}
 \includegraphics[scale=0.38]{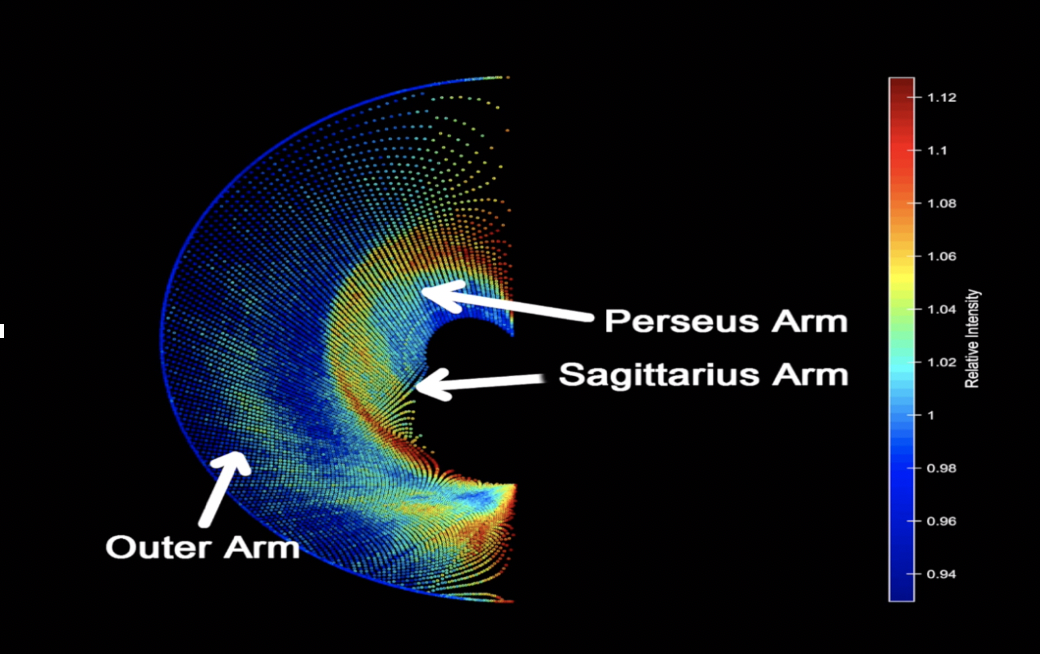}
\caption{Sections of Spiral Arms in Our Very Own Galaxy,(Ben-David, 2015).}
\label{Figure 7}
\end{figure}

\begin{figure}
 \includegraphics[scale=0.53]{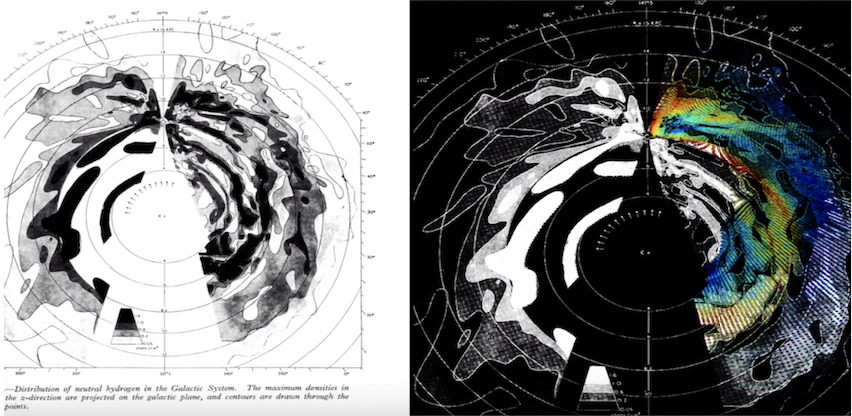}
\caption{The Map of the Milky Way Galaxy,(Oort, Westerhout,1958)).}
\label{Figure 8}
\end{figure}
\section{Arms of the Milky Way }
According to a study that utilized WISE data, Sagittarius, Perseus and Outer have been found to contain numerous star s embedded in clusters (NASA/JPL, 2017). The stars clusters are essential in understanding the formation and fate of the spiral arms. This is because the clusters contain the relatively younger stars whose process of formation is easy to follow. After formation, the new stars spend some time in the galaxy’s arms where there is a rich supply of gas. Later on stars move away from the arms. As such, the mapping of the galaxy relies heavily on the embedded structure as they play a complementary role, providing confirmatory results to what other studies such as radio telescope have found. Movement in the spiral arms is like traffic on a busy road. Just like vehicles in the jam move slowly in the jam, the star crowd and gas move together. The dense spiral arms compresses material passing through them, triggering a new process of star formations.
The embedded star clusters if better observed through WISE which uses Infra-Red light which not only shrouds the sky but also penetrates the thick dust in the galaxy. WISE gave a more elaborate image of the Infra-Red radiations cut across the dust that obstructs this clusters (Kormendy, Bender, 2018). WISE images were very effective due to the fact that the whole sky was scanned. In a Spitzer Space telescope mapping that counts the stars, the galaxy’s arms reveal higher concentration of stars than other regions. Just like the WISE. The Spitzer telescope uses Infra-Red rays.

\section{Formation of the Spiral structure}
Objects around the Milky Way galaxy do not rotate in a regular manner. Instead, the objects rotate in the fashion observed as observed in objects rotating around the solar system (Umen, 2018). The gas clouds, dust, and stars around the galaxy obey the law of planetary rotation. Objects closer to the center of the galaxy take shorter time to complete one rotation than objects further away from the center.
In general, interstellar objects and stars rotating at longer orbits naturally lag behind similar objects in smaller orbits in an effect referred to as differential galactic rotation. Seemingly, this effect explains the general observation a lot of material are concentrated in long arm-like spiral shaped features (Umen, 2018). Further, the differential rotation of the galaxy is thought to disrupt and disorient the original location of materials around the galaxy’s center, effectively rearranging them in a spiral pattern. The figure below shows the formation of the spiral arm. It is worth noting that objects further anyway from the galaxy’s center lag behind as the rotation.
\begin{figure}[h!]
 \includegraphics[scale=0.38]{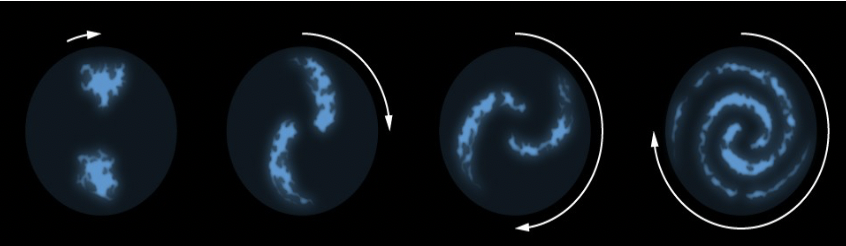}
\caption{Spiral Pattern, (Umen, 2018)}
\label{}
\end{figure}

At a glance, the theory may seem satisfactory in finding the solution to the puzzle posed by spiral formation as it conforms to logic and consistent with Kepler’s third law, but it immediately creates another puzzle. If such the spiral were formed due to differential rotation of objects around the center, then one would expect the disappearance of the spiral arms over the 13 billion years that the galaxy is thought to have existed. The differential rotation should wound the arms closer eventually merging them together into a single mass of a thin cylindrical disk. This speculation also raises questions as to whether the galaxy had the spiral arms when it was originally formed d whether such arms should last so long.
\par
To answer these questions, the discovery of the Hubble Space Telescope has played a vital role. The telescope can determine the structure of distant objects together with its appearance at their infancy stage more than 13 billion years ago. Information from the telescope reveals that galaxies are formed without spiral structure but have bright regions where stars are actively formed (Harrington, 2013).
\par
Generally, galaxies start to form a mature structure after several billion of existence and some lose their spiral structure. The galaxy then becomes less turbulence and rotation the main motion of both stars and gas. The disk in which stars form becomes quieter and the smaller clumps that form the stars form vaguely spiral. At later stage, after approximately 3.6 billion later, distant spiral formed. This was then followed by formation of the structure with several arms (Harrington, 2013). .
Through the calculation using supercomputer, scientist can investigate millions of particles that make up a given star to find out whether the effect of gravitational force can force them to form the spiral structure. Such calculations have revealed that gravitational effect of huge molecular effect is enough to cause spiral structure formation (Harrington, 2013). After formation, the arms can last for several billion years through a self-perpetuating system.  One aspect of the structure that is bound to change with time is its brightness. The change is occasioned by the commencement and completion of the process of star formation. Further, the arms are kept together by a gravitational force they exert.
 \par
 More recently, scientists have confirmed an “x” in the arrangement of stars in the bulge of the Milky Way. Using data from Wide-field infra-red explorer, WISE of NASA, one scientist Dustin Lang of Toronto University’s Dunlop institute took survey of the entire sky using infra-red radiation. Unlike the visible light, infra-red light can penetrate the dust in the galaxy, effectively allowing astronomers to observe details of the galaxy beyond the dust. When Lang eventually came up with a map of the galaxy and posted it on an interactive websites, other astronomers noted the strange shape of the galaxy’s bulge (Wegg, Gerhard, Portail, 2015).
After collaboration with an astronomical postdoctoral researcher at Max Plank Institute, Melisa Ness, a study was published to confirm the X-shaped pattern of stars on the galaxy’s bulge. Ness describes the “X” pattern as the galaxy’s signature and expressed its understanding as a big step towards understanding of the formation of Milky Way.

\begin{figure}[h!]
 \includegraphics[scale=0.50]{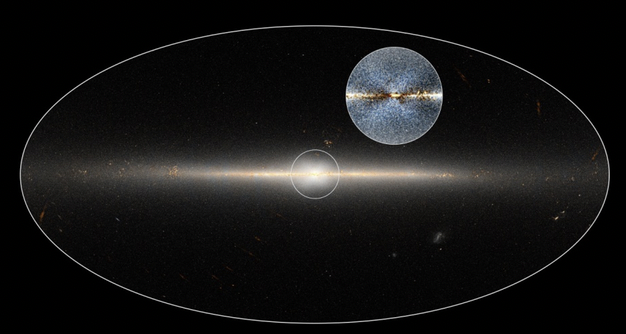}
\caption{The $X$-Shaped Pattern,(Wegg, Gerhard, Portail, 2015).}
\label{figure 14}
\end{figure}
According to the study, the “X” shaped bulge is observed  due to a lateral view of a group of stars revolving on their orbits and crossing the galactic center in a box-like distribution (Wegg, Gerhard, Portail, 2015). The bulge results when an unstable bars bucking in the center.  The study also explained the formation of the bar. It forms due to enlarging of a thin disk of star and gas. Since the Milky Way is one of the disk galaxies that have a bar, it is no surprise that $“X”$ shaped bar is observed.
Apart from Melissa and Lang’s publication many other studies have reported the $“X”$ shape of the galaxy’s bulge. For instance an experiment with the NASA’s COBE revealed a boxy-shaped bulge. In another study done in 2013 that included a 3D mapping o the Milky Way, the bulge had an $x$ shape but its exact shape was not shown (NASA, 2015).
 \section{The Galaxy’s Rotation Curve
}
One useful piece of information we would like to have about our galaxy it its rotation curve, which is a a graph of how fast things orbit in the galaxy  
as function of how far they are from its center. What might the Milky Way rotation curve look like? As mentioned above, the Milky Way galaxy is not stationary. It rotates around a center along with all stars and terrestrial bodies in it. Even then, it is important to establish Milky Way’s rotation curve, which relates the relative positions of objects around the center of the galaxy and the speed they move as shown in figure 11.

Determination of the actual rotation curve requires application of the Third Law of Planetary motion invented by Johannes Kepler (Kormendy, Bender, 2018). According to Kepler, the square of the time a body takes to complete an orbital rotation is directly proportional to the radius of rotation $T^2$ $\propto$ $r^3$
When the orbital velocity is introduced and the equation simplified further, it becomes;
$V$ $\propto$ ${\frac{1}{r}}^{\frac{1}{2}}$
This implies that the orbital speed is expected to be the inverse square root of the orbital radius. With such an expression, the expected rotational curve would appears as shown in figure 12.

\begin{figure}[h!]
 \includegraphics[scale=0.40]{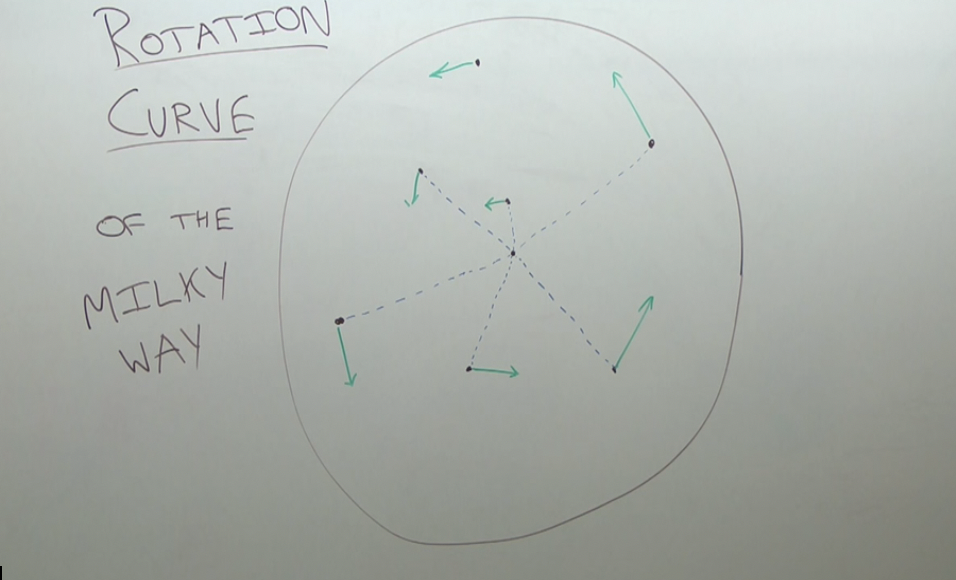}
\caption{ Milky Way’s Rotation curve,(Ben-David, 2015).}
\label{Figure 9}
\end{figure}

 \begin{figure}
 \includegraphics[scale=0.90]{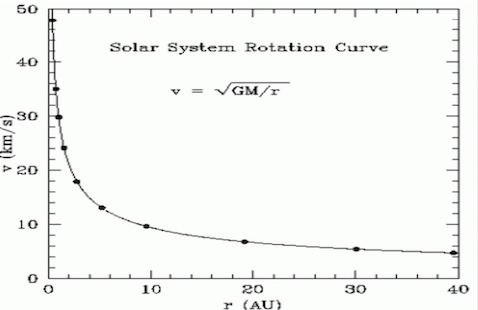}
\caption{ Solar System Rotation Curve, (Chen, 2017)}
\label{Figure 10}
\end{figure}
From the earlier discussion, it is possible to estimate the velocity of gas clouds using the spectrum produced from radio telescope signals (Kormendy, Bender, 2018). It is expected that the fastest gas cloud would be those closest to the center of the galaxy. As the various clouds are rotated by their orbital velocity around the center, one can then apply trigonometric ratios to locate exactly where they are, using the equation $y=x sin \theta$, where(see figure 13)

\begin{figure}[h!]
 \includegraphics[scale=1.5]{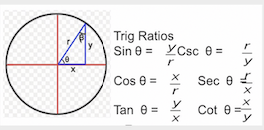}
\caption{ }
\label{}
\end{figure}

If the procedure is repeated for various gas clouds around the center of the galaxy, one can find the rotation curve of the galaxy. Then, the graph obtained is as follows in figure 14.\\
\newline
\begin{figure}[h!]
 \includegraphics[scale=0.55]{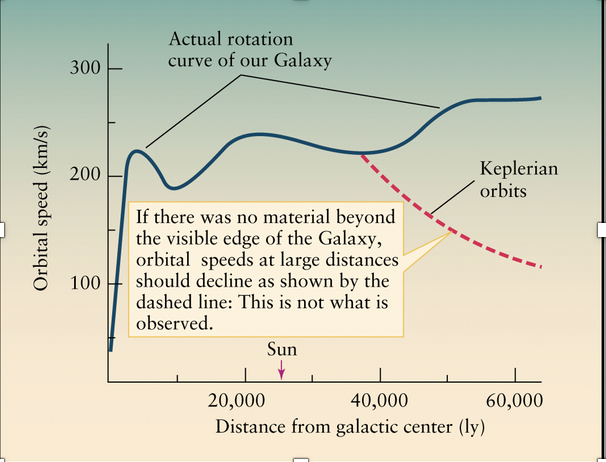}
\caption{The model of Kepler rotation curve and the observed one}
\label{}
\end{figure}
In comparison with Kepler’s prediction there is a big difference (Kormendy, Bender, 2018). Even with accurate such experiments, the actual rotation curve do not match the experimental one.  Moreover each time the experiment is repeated, the results are always pretty much the same. Was Kepler wrong? Not quite. We can also try flipping our reasoning and asking:- What would have to be different about our galaxy in order to see this kind of rotation curve? 
It turns out that you expect exactly this kind of curve if the galaxy has a lot more stuff in a particular arrangement. But when look out where this stuff should be, we do not see anything! This stuff does not interact with other normal matter, or with light, both which pass right through. But it does interact gravitationally and provides enough extra heft to the galaxy that things orbit at the speeds we observe. This indirect evidence for the existence of the mysterious  “Dark Matter”  whose composition and properties are an open problem in science right now. 
\section{Conclusion} \label{sec:conclusion}
Astronomers are getting closer to unmasking the mystery of structure of the Milky Way. The obstruction due to dust and concealed location of the earth on the galaxy may no longer be a hindrance to the mapping of the Milky Way thanks to radio telescope. The telescope has revealed that The Milky Way galaxy comprise of a disk that consist so stars with varying ages, mainly intermediate and young ones. The disk also contains clouds of gas that constantly give rise to new stars. The regions of active formation of new stars make up the spiral arms of the galaxy. It is good to note that radio observation of the Milky Way has undergone series of technological advancements such as development of Pan-STARRS1 that makes it possible to view even low mass brown and white dwarfs. The technology has high sensitivity, accuracy and muti-epoch observations making it possible to detect brown dwarfs, nearby stars, and stellar remnants for example white dwarfs found in north part of sky. Also, the distance and velocity of the objects can be measured relative to other stars and to determine multiplicity fraction by expanding neighboring sky to approximately 300 light years diameter. With this kind of technology in telescope, there has been increase in the number of brown dwarfs and variety of companions that cools dwarfs and to classification of low-mass star formed in stellar clusters. An example of discovered brown dwaff include PSO J318.5-22 that floats feely in the sky without orbiting a star and it is the lowest known brown dwarf. \\

%********************references***********************************

\end{document}